\documentclass[fleqn]{annalen}
\usepackage{amssymb}
\usepackage{graphicx}
\begin{document}
\newcommand{\volume}{}              
\newcommand{\xyear}{1999}            
\newcommand{\issue}{}               
\newcommand{\recdate}{29 July 1999}  
\newcommand{\revdate}{dd.mm.yyyy}    
\newcommand{\revnum}{0}              
\newcommand{\accdate}{dd.mm.yyyy}    
\newcommand{\coeditor}{ue}           
\newcommand{\firstpage}{1}         
\newcommand{\lastpage}{10}          
\setcounter{page}{\firstpage}        
\newcommand{\keywords}{dephasing, low-dimensional conductors,
electron-phonon interaction} 
\newcommand{\PACS}{72.15.Rn, 73.23.-b, 72.10.Di}
\newcommand{\shorttitle}{M. E. Gershenson, Low-temperature dephasing in 
disordered conductors} 
\title{Low-temperature dephasing in disordered conductors: experimental aspects}
\author{M. E. Gershenson} 
\newcommand{\address}
  {Serin Physics Laboratory, Rutgers University, Piscataway, NJ 08854-8019,
   USA}
\newcommand{\email}{\tt gersh@physics.rutgers.edu} 
\maketitle

\begin{abstract}
What is the lowest temperature to which one can trace the growth of the dephasing time $%
\tau_\varphi $ in low-dimensional conductors? I consider the fundamental
limitation, the crossover from weak to strong localization, as well as
several experimental reasons for frequently observed saturation of $%
\tau_\varphi $ (hot-electron effects, dephasing by external noise).
Recent progress in our understanding of the electron-phonon interaction in
disordered conductors is also briefly discussed.
\end{abstract}

\setcounter{page}{\firstpage} 




\section{Introduction}

Electron localization and dephasing of the electron wave function are
deeply interwoven. This idea was established more than two decades ago,
and is widely appreciated now \cite{aa,imry}. The quantum interference
measurements in the weak localization (WL) regime became a powerful tool for
the study of different scattering processes in disordered conductors. As a
result of literally hundreds of experiments, the dephasing time $\tau
_{\varphi }$ has been measured in all dimensions. The earlier progress in
this field was summarized in reviews \cite{bergmann, aags}; more recent
results can be found in Refs. \cite{imry, bouchiat}.

The dephasing attracted renewed interest due to rapid development of the
quantum information processing (see, e.g. \cite{vinc}). The relatively short
dephasing time in solid-state devices remains one of the major obstacles
in the way of implementation of quantum computers. The importance of universal
(intrinsic) and non-universal (extrinsic) limitations on the dephasing rate
has been emphasized in a recent paper by Mohanty, Jarivala, and Webb \cite
{mohanty}. The authors of Ref. \cite{mohanty} addressed the old problem of the
low-temperature saturation of $\tau _{\varphi }$, which is observed in
essentially all experiments at $T<0.1-1K$. On the basis of detailed
measurements of $\tau _{\varphi }$ in one-dimensional ($1d$) gold wires,
they came to the conclusion that a finite dephasing rate at $T=0$ is
intrinsic, and is related to the zero-point fluctuations of the electrons.
This issue is important since an intrinsic saturation of $\tau _{\varphi }$
would signal a breakdown of Fermi-liquid behavior of low-dimensional
conductors in the metallic regime.

This mini-review does not pretend to cover all recent measurements of $\tau
_{\varphi }$. Instead, I focus on several problems, \-which are common for
the low-temperature dephasing experiments, and which are relevant to the
quest for extra-large $\tau _{\varphi }$. One problem is fundamental: when $%
\tau _{\varphi }$ becomes of the order of the diffusion time over the
localization length $\xi $, the low-dimensional conductors enter the strong
localization (SL) regime. In this regime, the hopping rate replaces the
dephasing rate. Two other experimental problems are the noise-induced
dephasing and overheating of electrons by applied voltage and/or external
noise. Recent measurements of the electron-phonon scattering rate in
disordered conductors help to clarify the latter problem.

\section{Sources of dephasing}

The dephasing time is usually extracted from the low-field magnetoresistance 
\cite{bergmann,aags} and, less often, from the temperature dependence of the
universal conductance fluctuations \cite{webb,birge}. Measurements of $\tau
_{\varphi }$ in a variety of 1d, 2d, and 3d metal and semiconductor
stuctures have been reported (see \cite
{bergmann,aags,bouchiat,polyan,dugdale} and references therein). At present,
there is experimental evidence of several ''intrinsic'' and ''extrinsic''
sources of dephasing. The most important intrinsic mechanisms are:

- the electron-phonon interaction, which governs dephasing at high
temperatures\footnote{%
In 3d conductors, this temperature range can be extended down to 
sub-Kelvin temperatures (see Fig. 5b).};

-the electron-electron interaction, which dominates at low temperatures
(typically, below 1-10K for 1d and 2d conductors).

The most common ''extrinsic'' sources of dephasing are magnetic impurities
and the high-frequency electromagnetic noise in the experimental set-up.

I will focus on the low-$T$ regime in 1d and 2d conductors, typical
experimental data are shown in Fig.1. The observed
dependence $\tau _{\varphi }(T)$ can be accounted for by the theory of
electron-electron interaction in disordered metals \cite{aa,aak}. The theory
predicts that in low dimensions, interactions with small momentum and
energy transfers are the most efficient dephasing mechanism. These
quasielastic electron-electron collisions are equivalent to interaction of
an electron with the fluctuating electromagnetic field produced by all the other
electrons (dephasing by the equilibrium Nyquist noise). The Nyquist
dephasing rate can be expressed as \cite{aak}:

\begin{equation}
1d:\hbox{ \ \ \ \ }\tau _{\varphi }^{-1}=\left( \frac{e^{2}\sqrt{2D}k_{B}TR}{%
\hbar ^{2}L}\right) ^{2/3},  \label{1dee}
\end{equation}

\begin{equation}
2d:\hbox{ \ \ \ \ }\tau _{\varphi }^{-1}=k_{B}T\cdot \frac{e^{2}R_{\square }%
}{2\pi \hbar ^{2}}\ln \left( \frac{\pi \hbar }{e^{2}R_{\square }}\right)
^{2/3}.  \label{2dee}
\end{equation}
\begin{figure}[th]
\begin{center}
\includegraphics[angle=270, width=4in]{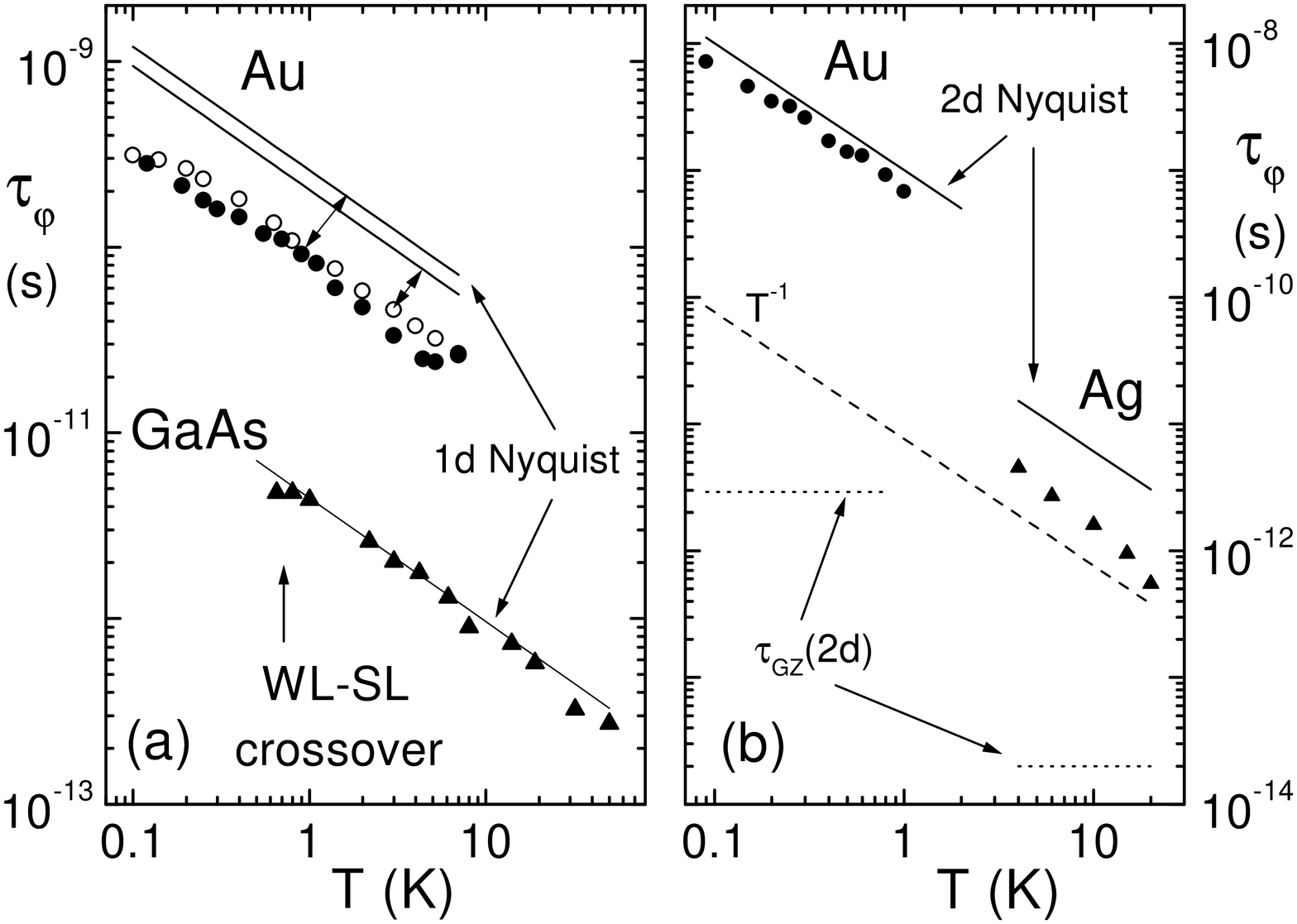}
\caption{(a) The dependences $\protect\tau _{\protect\phi }(T)$ for Au wires ($%
\circ $ - the width $W=90nm$, $R_{\square }=3\Omega $, $D=135cm^{2}/s$; $%
\bullet $ - $W=60nm$, $R_{\square }=3.1\Omega $, $D=116cm^{2}/s$) 
\protect\cite{echt-ee}, and for the Si $\protect\delta $-doped GaAs wire ($%
\blacktriangle $ - $W=50nm$ , the carrier concentration $3\cdot
10^{12}cm^{-2}$, and $R_{\square }=1k\Omega $) \protect\cite{prl2-gersh}. The
solid lines are the 1d Nyquist $\protect\tau _{\protect\varphi }(T)$ 
(Eq.\ref{1dee}). In the $\protect\delta $-doped
wire, the electrons are strongly localized below $T\simeq 0.4K$. For the Au
wires, the WL-SL crossover is expected at $\protect\mu K$ temperatures, however,
the noise-induced saturation of $R(T)$ and $\protect\tau _{\protect\varphi }(T)$
was observed below $0.1K$. (b) The dependences $\protect\tau _{\protect\phi }(T)$ for Au 
films ($\bullet $ -  $R_{\square }=32.7\Omega $, $\tau =1.8 \cdot 10^{-15}s$,
$k_{F}l=31$) \protect\cite{pierre},
and for ultrathin epitaxial Ag films ($\blacktriangle $ -  $R_{\square }=1.5k\Omega $, 
$\tau =5.8 \cdot 10^{-16}s$, $k_{F}l=10$) \protect\cite{henzler}. The solid lines 
are the 2d Nyquist $\protect\tau _{\protect\varphi }(T)$ (Eq.\ref{2dee}), the dotted 
lines - the maximum $\tau _{GZ}(2d)$ \protect\cite{gz}.} 
\label{1dAu}
\end{center}
\end{figure}
Here $D$ is the electron diffusion constant, $R_{\square }$ is the sheet
resistance, and $R/L$ is the resistance of a 1d conductor per unit length. The
predicted temperature dependences of the dephasing rate [$\tau _{\varphi
}^{-1}(1d)\propto T^{2/3}$,  $\tau _{\varphi }^{-1}(2d)\propto T$] agree
very well with the experimental data for many systems, including thin metal
films and wires 
\cite{echt-ee,prl2-gersh,pierre,henzler,lin,krasovitskii,wind,thornton}, 
semiconductor structures \cite{tsui,katine}, and even multi-wall carbon 
nanotubes \cite{nanotubes}.
Absolute values of $\tau _{\varphi }$ for metal wires are usually somehow
smaller than the estimates (\ref{1dee}) and (\ref{2dee}) \footnote{%
Note that for correct extraction of $\tau _{\varphi }$ from the 1d
magnetoresistance, one should use the expression for the 1d WL correction
calculated for the quasi-elastic dephasing in Ref. \cite{aak}. Use of the WL
correction for the strongly-inelastic processes [B. L. Altshuler and A. G.
Aronov, Sov. Phys. JETP\ Lett. \textbf{33} (1981) 515] would result in
overestimation of $\tau _{\varphi }$ by a factor of 4 at low $T$, where the
Nyquist mechanism dominates \cite{echt-ee}.}(see Figs. 1, 2a, 3b, and 5a).  

Both expressions (\ref{1dee}) and (\ref{2dee}) can be combined as \cite{aa,aga}:

\begin{equation}
\frac{\hbar }{\tau _{\varphi }}\hbox{\ \ }\sim \frac{k_{B}T}{g(L_{\varphi })}
\label{genee}
\end{equation}
where $g(L_{\varphi })\varpropto L_{\varphi }^{d-2}$ is the conductance of a 
$d$-dimensional system (measured in units of $e^{2}/h$) on the scale
determined by the phase-breaking length $L_{\varphi }=\sqrt{D\tau _{\varphi }%
}$. On the ''metallic'' side of the
WL-SL crossover, where $g(L_{\varphi })\ll g(\xi )\sim 1$, the electron states are
well-defined ($T\tau _{\varphi }>>1$). Thus, both 1d and 2d conductors
should behave as Fermi-liquid systems over the whole $T$ range that
corresponds to the WL regime \cite{quasi-1d}. At the same time, Eq. \ref
{genee} shows that the WL-SL crossover imposes a fundamental
limitation on the increase of the dephasing time: $\tau _{\varphi }$
cannot exceed the diffusion time over the localization length. Recent
experiments on observation of the Thouless crossover in 1d conductors
confirm this prediction \cite{prl2-gersh,prl1-gersh,prb-gersh} (see Fig.
2). All relevant energy scales in 1d conductors become of the same order at
the crossover temperature $T_{\xi }$ \cite{prb-gersh}:

\begin{equation}
\frac{\hbar }{\tau _{\varphi }(T_{\xi })}\hbox{\ }\sim k_{B}T_{\xi }\hbox{\ }%
\sim \Delta _{\xi }\sim \frac{\hbar D}{\xi ^{2}}\sim k_{B}T_{0}
\label{allenergy}
\end{equation}
Here $\Delta _{\xi }=(\nu _{1}\xi )^{-1}$ is the mean level spacing within
the localization domain ($\nu _{1}$ is the 1d single-particle density of
states), $\frac{\hbar D}{\xi ^{2}}$ is the Thouless energy, and $T_{0}$ is
the activation energy of the Arrhenius dependence $R(T)$ observed in the 
SL regime.

\begin{figure}[th]
\begin{center}
\includegraphics[angle=270, width=4in]{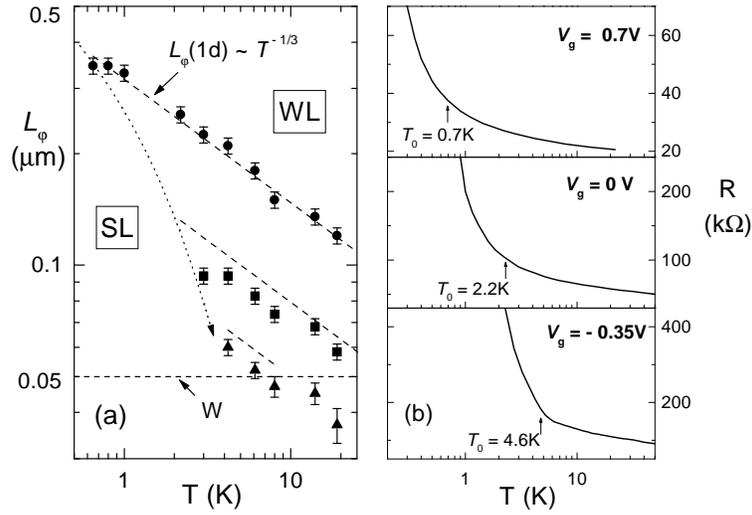}
\caption{(a) The dependences $\protect\tau _{\protect\varphi }(T)$ for 1d 
\textit{Si} $\protect\delta $-doped \textit{GaAs} wires ($360$ wires in
parallel, each wire $0.05\protect\mu m$ wide and $500\protect\mu m$ long) at
three values of the gate voltage $V_{g}$ : $0.7V$ ($\bullet $), $0V$ 
($\blacksquare $), $-0.35V$ ($\blacktriangle $). (b) The dependences $R(T)$
for these values of $V_{g}$. The arrows show the activation
energy $T_{0}$ of the Arrhenius dependence $R(T)$ observed in the SL regime 
\protect\cite{prl2-gersh}.}
\label{1dWLSL}
\end{center}
\end{figure}

Demonstration of this universality is facilitated by the ''sharpness'' of
the temperature-driven WL-SL crossover in 1d conductors.
It is more difficult to access both the WL and SL regimes for the same 2d 
sample without changing its parameters.

\begin{figure}[th]
\begin{center}
\includegraphics[angle=270, width=3.7in]{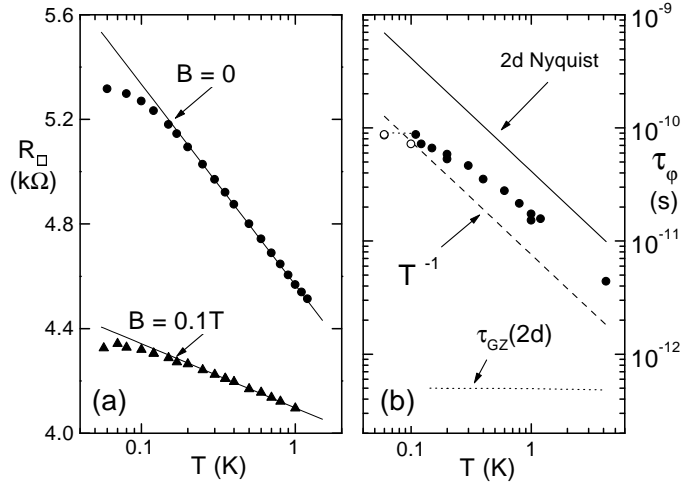}
\caption{(a) The temperature dependences of the sheet resistance for 2d 
\textit{Si} $\protect\delta -$doped \textit{GaAs} ($n=1\cdot 10^{12}cm^{-2}$,
$\tau = 4.8\cdot 10^{-14}s$, $k_{F}l=5.2$) 
at $B=0$ ($\bullet $) and at $B=0.1T$ 
($\blacktriangle $) \cite{gkrsw}. The solid lines are the
logarithmic fit $R_{\square }(T)=R_{\square }(1K)\left[ 1-\protect\alpha 
\frac{e^{2}R_{\square }}{2\protect\pi ^{2}\hbar }\ln \left( T/1K\right) %
\right] $ with $\protect\alpha =1.29$ ($B=0$) and $\protect\alpha =0.52$
($B=0.1T$). (b) The dependence $\protect\tau _{\protect\varphi }(T)$ for
this sample ($\circ $ - without correction for the electron
overheating). The solid line is the 2d Nyquist dephasing time (Eq.\ref
{2dee}), the dotted line - $\tau_{GZ}(2d) $ \cite{gz}.}
\label{2dgaas}
\end{center}
\end{figure}

\section{Low-temperature saturation of $\protect\tau _{\protect\varphi }$}

The crossover temperature is usually unattainably low for thin metal wires
and films (unless their thickness is comparable with the Fermi wavelength 
\cite{goldman,hsu}). Nevertheless, all experiments with disordered metals
and semiconductors  show some saturation of $\tau _{\varphi }$ at $T<0.1-1K$
(for references, see \cite{mohanty}). There are several possible causes for
saturation of $\tau _{\varphi }$, the most common ones are scattering by
localized spins and overheating of the electrons (due to the applied current
or to the external noise). A more subtle reason for saturation is 
noise-induced dephasing without overheating of electrons.

\subsection{''Hot'' electrons and the electron-phonon interaction}

The first problem is related to the electron-phonon interaction (EPI). At
low temperatures, the EPI is the bottleneck in the energy transfer from
the electrons to the heat sink, and the electrons can be easily overheated
by the measuring current or external noise (another mechanism of heat
removal, the heat flow along the sample in cooler leads, will be considered
below). Thus, the knowledge of the EPI is essential for the low-temperature
dephasing measurements.

\begin{figure}[th]
\begin{center}
\includegraphics[angle=270, width=3.7in]{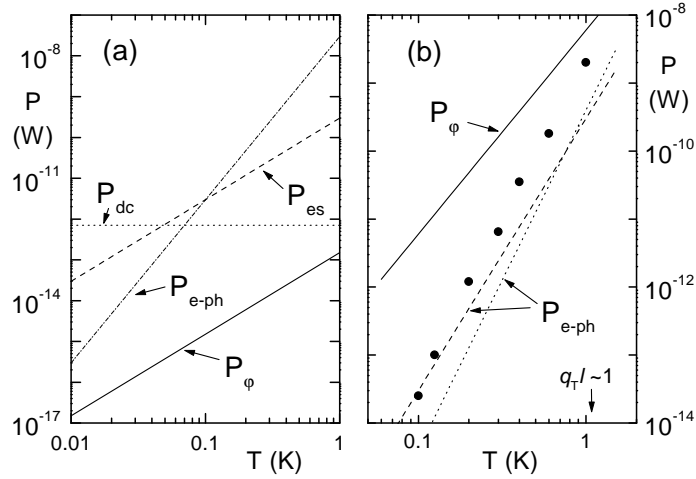}
\caption{(a) The temperature dependences of $P_{\protect\varphi }$ (the rf
noise power sufficient for saturation of $\protect\tau _{\protect\varphi }(T)
$), $P_{e-ph}$ (the heat flow from the electrons to phonons),
$P_{es}$ (the heat flow due to outdiffusion of the hot
electrons in cooler leads), calculated at $T_{e}=1.3T$ and $T_{ph}=T$
for sample Au-2 from Ref. \protect\cite{mohanty} 
(for details of calculation, see \protect\cite{aga}).
The horizontal dotted line is the bias current power. (b) The
bias current power, which corresponds to an increase of the
electron temperature $T_{e}$ by $10\%$ over the mixing chamber temperature $T$
for a 2d\textit{\ Si} $\protect\delta -$doped \textit{GaAs} device with the
area $0.36$x$0.12mm^{2}$ (the other parameters - as in Fig.3) \protect\cite
{gkrsw}. The dependences $P_{e-ph}$ due to the piezoelectric coupling in GaAs are
calculated for $T_{e}=1.1T$: the dashed line - for $q_{t}l<1$ ($q_{t}$ is
the wave vector of a thermal phonon, $l$ is the electron mean free path) 
\protect\cite{girvin}, the dotted line - for $q_{t}l>1$ \protect\cite{price}%
. }
\label{epi}
\end{center}
\end{figure}
The last 15 years have resulted in considerable progress in understanding of the
EPI in disordered metals and heavily doped semiconductors. It has been
predicted theoretically \cite{rammer,reizer} and tested experimentally \cite
{echt-ph,ptitsina1,ptitsina2}, that the interactions with \textit{transverse} 
phonons are dominant at $q_{T}l<1$ ($q_{T}$ is the wave number of a
thermal phonon, $l$ is the electron mean free path) (for a review, see \cite
{rs}). Notice that in the clean case, the electrons interact with
longitudinal phonons only. The constants of interaction with transverse
phonons can be found from analysis of the $T^{2}$ term in the dependence $%
R(T)$ for disordered metals \cite{reizer,echt-ph,ptitsina2}. Using the
interaction constants, one can estimate the contributions of transverse and
longitudinal phonons to the electron-phonon scattering time $\tau _{e-ph}$.
These calculations are in good agreement with direct measurements of $\tau
_{e-ph}$ in thin metal films \cite{ptitsina2} (see also Fig.5b). 

Rapid growth of $\tau _{e-ph}$ with decreasing temperature causes a very
large thermal resistance between the electrons and the phonons. Figure 4a
shows that a sub-pico-watt power $P$ can overheat electrons
in a 1d wire at $T=0.1K$. Even smaller values of $P$ are
sufficient for overheating the 2d electrons in GaAs structures (Fig. 4b).
For this reason, a very thorough filtering of all the leads to the sample is
required.

Since both localization and interaction corrections at low $T$ depend on the
electron temperature only, it is easy to check if the electrons are at the
bath temperature: the electron overheating also causes saturation of  
$R(T)$. Figure 3 shows $R(T)$ for 2d Si $\delta $-doped GaAs structure at $%
B=0$ and at $B=0.1T$ (the latter magnetic field is sufficiently strong to
suppress the temperature dependence of the WL correction) \cite{gkrsw}.
Saturation of the logarithmic dependence $R(T)$ indicates that the electrons
were never cooled below 0.1K.
The dephasing time plotted for this sample versus the bath temperature $%
T$ demonstrates a clear saturation below $T=0.15K$ (Fig 3); this saturation
is less pronounced if one takes into account the electron overheating.

\subsection{The noise-induced dephasing}

The authors of Ref.\cite{mohanty} have demonstrated that saturation of 
$\tau _{\varphi }$, observed for 1d Au wires, was not caused by electron 
overheating or by the spin-spin scattering \footnote{%
The authors of Ref.\cite{mohanty} ruled out the possibility of conventional
spin-flip scattering. However, it has been suggested recently that dephasing
by two-level systems (TLS) in the two-channel Kondo regime can cause
saturation of $\tau _{\varphi }(T)$ in a limited temperature range \cite
{zawad}. This mechanism is non-universal, since the density of the
non-magnetic dynamic TLS with an almost degenerate Kondo ground state is both
sample- and history-dependent.}. It is more difficult, however, to rule out
another possibility for the $\tau _{\varphi }(T)$ saturation, namely, 
dephasing by external noise. Indeed, it has been known since the early 80s,
that the external high-frequency radiation can induce dephasing \cite
{aak1}; this effect was observed in the experiments \cite
{wang,vitkalov,giordano}. The key point is that dephasing by the rf
noise can occur \textit{without} electron overheating. 

What are the experimental conditions for observation of this effect? The
phase coherence is destroyed most efficiently by spectral components of the
noise with the frequency $\omega \sim \tau _{\varphi }^{-1}$ \cite{aak1}. The
electric field $E_{\varphi }$ for dephasing at the time scale $\tau
_{\varphi }$ can be estimated as $E_{\varphi }\approx \hbar /eL_{\varphi
}\tau _{\varphi }\propto \tau _{\varphi }^{-3/2}$ \cite{aak1}. For 1d
conductors, $E_{\varphi }\propto $ $T$ (see Eq.\ref{1dee}), and the noise
power sufficient for saturation of $\tau _{\varphi }(T)$ below a certain $T$
is \cite{prl2-gersh,aga}: 
\begin{equation}
P_{\varphi }=\frac{(LE_{\varphi })^{2}}{2R}=R\left( \frac{ek_{B}T}{\hbar }%
\right) ^{2}.
\end{equation}
The inflow of the noise energy is balanced by the energy outflow due to
two competing cooling mechanisms, the EPI and outdiffusion of hot 
electrons in cooler leads. The latter mechanism
becomes more efficient in relatively short samples at low
temperatures \cite{prober,mittal}. As it is shown in Refs. \cite
{prl2-gersh,aga}, balancing of the incoming rf noise energy (sufficiently
strong for dephasing) by the hot-electron outdiffusion  results in a
negligible rise of the electron temperature if the total resistance of a
wire is less than the quantum resistance $R_{Q}=h/e^{2}=25.8k\Omega $. In
the experiment \cite{mohanty}, the wire resistance was in the range $%
R=0.3-3.6k\Omega $ , and one cannot exclude the possibility of 
rf-noise-induced dephasing without electron overheating
(see Fig. 4a). Experiments with wires of
a larger $R>>R_{Q}$ could clarify this situation. Such high-resistance
wires are also less prone to the noise-induced dephasing, since $P_{\varphi }
$ is proportional to $R$. However, there is a trade-off: longer wires cannot
be efficiently cooled by the hot-electron outdiffusion, and restrictions on
the measuring current become more severe.

\begin{figure}[th]
\begin{center}
\includegraphics[angle=270, width=4in]{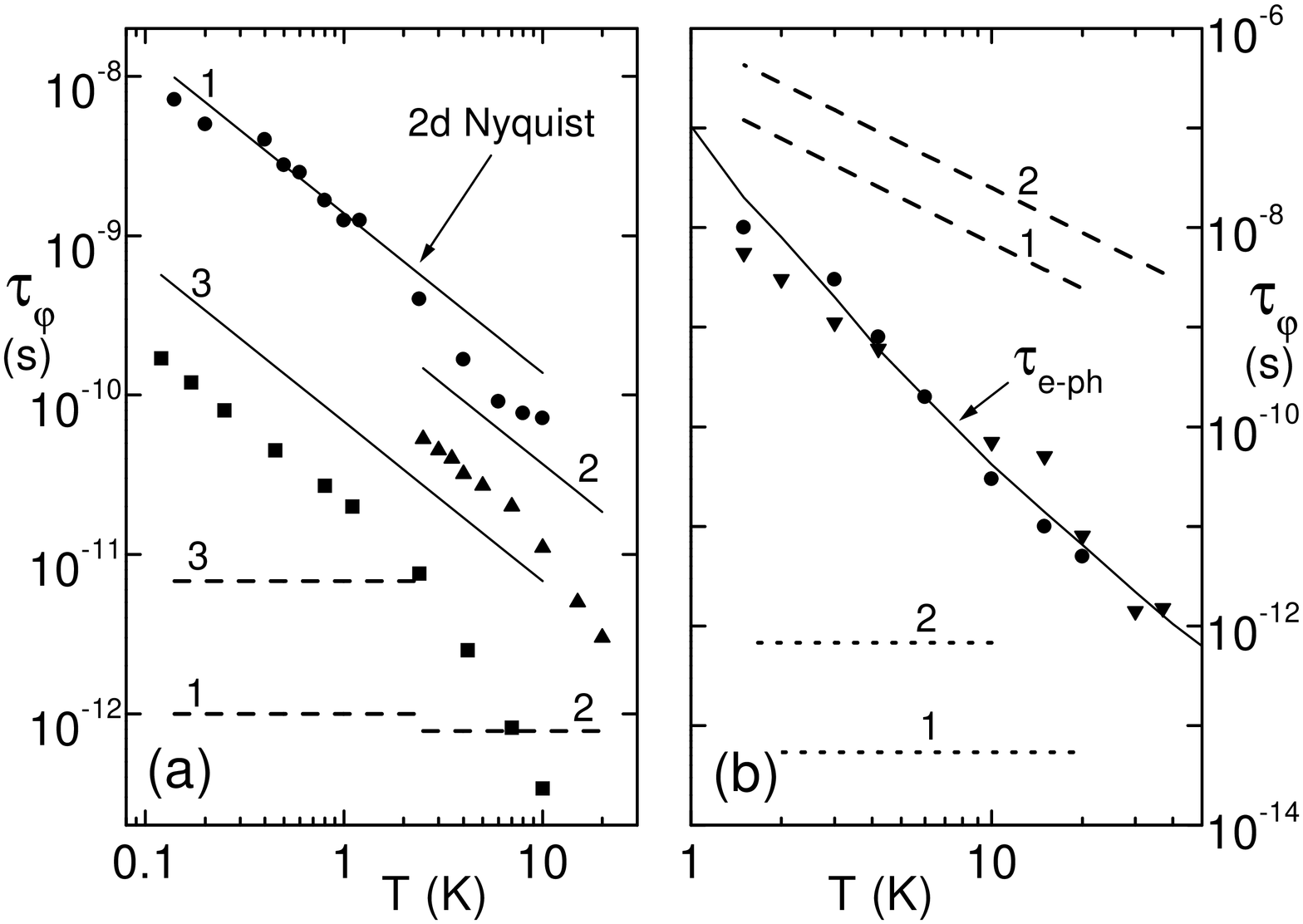}
\caption{(a) The dependences $\protect\tau _{\protect\varphi }(T)$ for 2d
metal films: 1 ($\bullet $)\ - Mg film with $R_{\square }=22.3\Omega $, 
$\protect\tau =4.6\cdot 10^{-16}s$, $k_{F}l=10$  
\protect\cite{white}; 2 ($\blacktriangle $) - Al film with 
$R_{\square }=112\Omega $, $\protect\tau =1.7\cdot 10^{-15}s$,
$k_{F}l=60$  \protect\cite{ggz}; 3 ($\blacksquare $) - Bi film with 
$R_{\square }=630\Omega $, $\protect\tau =2.8\cdot 10^{-14}s$ 
\protect\cite{komori}. (b) The dependences $\protect\tau _{\protect\varphi
}(T)$ for 3d conductors: $\bullet $ - $1\protect\mu m$-thick disordered Cu
films with $\protect\rho =6\cdot 10^{-5}\Omega \cdot cm$, $%
\protect\tau =5\cdot 10^{-16}s$, $k_{F}l=11$  \protect\cite{agz}; $\blacktriangledown $ -
solid solution $Cu_{0.9}Ge_{0.1}$ with $\protect\rho =2.8\cdot 10^{-5}\Omega
\cdot cm$, $\protect\tau =1\cdot 10^{-15}s$, $k_{F}l=24$ 
\protect\cite{gey}. The solid line - $\protect\tau _{e-ph}$ calculated for Cu with $%
D=10cm^{2}/s$, the transverse-phonon contribution is greater by a factor of
40 than the longitudinal-phonon one at $T=2K$
(for details of calculation, see \protect\cite{ptitsina2}. The dashed lines - the
3d dephasing time due to the electron-electron inteactions \protect\cite
{aak}. The dotted lines - the maximum $\protect\tau _{GZ}(3d)$ \protect\cite{gz}.}
\label{2d3d}
\end{center}
\end{figure}

\section{Theory of dephasing by quantum fluctuations: comparison with experiment}

In an attempt to explain the low-$T$ saturation of $\tau _{\varphi }$, it has
been suggested that the zero-point fluctuations of electrons contribute to
dephasing \cite{mohanty}. Golubev and Zaikin obtained the finite
dephasing rate at $T=0$ in all dimensions due to quantum fluctuations of 
the electric field produced by other electrons \cite{gz}. 
The detailed criticism of this
idea, in general, and of the calculations by Golubev and Zaikin, in particular,
have been given in Refs. \cite{aga,aag}; here we briefly compare the
predictions of the theory \cite{gz} with the experiment.

According to Ref.\cite{gz}, the dephasing rate in 3d conductors is
limited by $\tau _{GZ}(3d)=\left( \frac{3\pi ^{2}\hbar \sigma }{e^{2}}%
\right) \sqrt{2D}\tau ^{3/2}$, which is comparable with the elastic
scattering time $\tau \sim 10^{-16}-10^{-14}s$. If this were the case, the
3d WL corrections would be completely ''washed out''. Nevertheless, the
temperature-dependent WL corrections have been observed in a host of 3d systems
(disordered metals \cite{aags}, metal glasses \cite{dugdale}, heavily doped
semiconductors \cite{polyan}). Figure 5b shows that dephasing 
in 3d metal films is governed by interaction with transverse
phonons. The experimental values of $\tau _{\varphi }$ exceed
the limiting value $\tau_{GZ}(3d)$ by 5(!) orders of magnitude. 
This disagreement is equally
strong for both macroscopically homogemeous and inhomogeneous systems
(thus, it cannot be attributed to the charging effects \cite{gz}),
with the parameter $k_{F}l>>1$ (this rules out the argument that the
theory \cite{gz} disagrees only with the data for systems with $k_{F}l \sim 1$).

Similar disagreement is observed in 2d. On the one hand, the upper limit $%
\tau _{GZ}(2d)=\left( \frac{4\pi \hbar }{e^{2}R_{\square }}\right) \tau $ would
preclude observation of the $T$-dependent WL corrections. On the other hand,
a well-pronounced temperature dependence of $\tau _{\varphi }$ has been
observed, and the experimental values of $\tau _{\varphi }$ are typically
much greater than $\tau _{GZ}(2d)$ (Figs. 1b, 3b and 5a). Note, that 
since $\sqrt{D\tau _{GZ}(2d)}$ is always much smaller than the localization
length, this would prohibit the 2d WL-SL crossover. 
This prediction is also at odds with the experiment \cite
{goldman,hsu,dahm,gkrsw}.

In 1d, the temperature dependence of $\tau_{GZ}^{-1}$ coincides with the 
1d Nyquist dephasing rate (Eq.\ref{1dee}) at 
$T\gg \hbar /\sqrt{\tau _{GZ}(1d)\tau }$,
and saturates at lower temperatures \cite{gz}. This saturation corresponds to the maximum
dephasing length $L_{\varphi }\approx 0.6\xi /\sqrt {N}$, where $\xi $ is 
the 1d localization length, $N$ is the number of transverse channels
in the wire. If this were the case, the 1d conductors would not enter the strong 
localization regime with decreasing temperature. However, the Thouless crossover 
has been observed for 1d conductors with $N=7-30$ \cite{prl2-gersh,prl1-gersh,prb-gersh}.  

\section{Conclusions}

At present, there is no experimental evidence for an
''intrinsic'', universal mechanism of a finite dephasing rate at $T=0$. Such
evidence could be obtained only if all the known reasons for saturation
of  $\tau _{\varphi }$ were ruled out. In this respect, a useful guideline is
provided by the Law of Economy (also known as Ockham's razor): ''entities
are not to be multiplied beyond necessity''. This does not mean, of course,
that one cannot expect to encounter new  phenomena with a further decrease in 
temperature. Clearly, the experimental and theoretical study of 
the low-temperature dephasing will remain an important and exciting 
field in the future.  

\vspace*{0.25cm} \baselineskip=10pt{\small \noindent I thank Yu. Khavin, 
P. Echternach, A. Bogdanov, D. Reuter, 
P. Schafmeister, and A. Wieck
for collaboration, and B. Altshuler and I. Aleiner for numerous helpful
discussions. This work is supported in part by the ARO-administered MURI
grant DAAD 19-99-1-0252.}


\begin{thebibliography}{99}
\bibitem{quasi-1d}  There are many transverse quantum channels in the
conductors that are one-dimensional with respect to the quantum interference
effects, and the inter-channel scattering is strong. The physics of these
(quasi)-1d conductors is essentially different from the physics of quantum
wires; the latter single-channel limit is described by the
Tomonaga-Luttinger model.
\end{thebibliography}
\end{document}